\documentclass[12pt, preprint]{emulateapj}
\usepackage{apjfonts}
\usepackage{soul}
\usepackage{color}
\usepackage{graphicx} 
\newcommand{\HI}{\textsc{Hi}}
\newcommand{\HII}{\textsc{Hii}}
\newcommand{\masq}{\mbox{mag~arcsec$^{-2}$}}
\def\aj{AJ}
\def\apj{ApJ}
\def\apjl{ApJ}
\def\apjs{ApJS}
\def\aap{A\&A}
\def\mnras{MNRAS}
\def\nat{Nature}
\def\pasp{PASP}%

\shortauthors{{Paudel et al.}}
\shorttitle{Destruction of dwarf galaxies around NGC 4216}

\begin{document}

\title{The Next Generation Virgo Cluster Survey. IV. NGC 4216: A Bombarded Spiral in the Virgo Cluster\altaffilmark{*}}
\altaffiltext{*}{Based on observations obtained with MegaPrime/MegaCam, a joint project of CFHT and CEA/DAPNIA, at the Canada-France-Hawaii Telescope (CFHT) which is operated by the National Research Council (NRC) of Canada, the Institut National des Sciences de l'Univers of the Centre National de la Recherche Scientifique of France and the University of Hawaii.}
\author{Sanjaya Paudel\altaffilmark{1$\dagger$},
Pierre-Alain Duc\altaffilmark{1},
Patrick C\^ot\'e\altaffilmark{2},
Jean-Charles Cuillandre\altaffilmark{3},
Laura Ferrarese\altaffilmark{2},
Etienne Ferriere\altaffilmark{1},
Stephen D. J. Gwyn\altaffilmark{2},
J. Christopher Mihos\altaffilmark{4}, 
Bernd Vollmer\altaffilmark{5}, 
Michael L. Balogh\altaffilmark{6}, 
Ray G. Carlberg\altaffilmark{7}, 
Samuel Boissier\altaffilmark{8}, 
Alessandro Boselli\altaffilmark{8},
Patrick R. Durrell\altaffilmark{9}, 
Eric Emsellem\altaffilmark{10\&11}, 
Lauren A. MacArthur\altaffilmark{2},
Simona Mei \altaffilmark{12}, 
Leo Michel-Dansac \altaffilmark{10},
Wim van Driel\altaffilmark{12},
}
\affil{
$^1$ Laboratoire AIM Paris-Saclay, CNRS/INSU, Universit\'e Paris Diderot, CEA/IRFU/SAp, 91191 Gif-sur-Yvette Cedex, France\\
$^2$National Research Council of Canada, Victoria, BC, V9E 2E7, Canada\\
$^3$Canada-France-Hawaii Telescope Corporation, Kamuela, HI 96743, USA\\
$^4$Department of Astronomy, Case Western Reserve University, 10900 Euclid Avenue, Cleveland, OH 44106, USA\\
$^5$Observatoire Astronomique, Universit\'e de Strasbourg \& CNRS UMR 7550, 11 rue de l'Universit\'e, 67000 Strasbourg, France\\
$^6$Department of Physics and Astronomy, University of Waterloo, Waterloo, Ontario, N2L 3G1, Canada\\
$^{7}$Department of Astronomy and Astrophysics, University of Toronto, Toronto, ON M5S 3H4, Canada\\
$^8$Aix Marseille Universit\'e, CNRS, LAM (Laboratoire d'Astrophysique de Marseille) UMR 7326, 13388 Marseille, France\\
$^9$Department of Physics and Astronomy, Youngstown State University, One University Plaza, Youngstown, OH 44555, USA\\
$^{10}$ Universit\'e de Lyon 1, CRAL, Observatoire de Lyon, 9 av. Charles Andr\'e, 69230 Saint-Genis Laval; CNRS, UMR 5574; ENS de Lyon, France\\
$^{11}$ European Southern Observatory, Karl-Schwarzchild-Str. 2, D-85748 Garching, Germany\\
$^{12}$GEPI,  Observatoire de Paris, CNRS, Universit\'e Paris Diderot, 5 Place J. Janssen, 92190 Meudon Cedex, France\\
}
\altaffiltext{$\dagger$}{Email: sanjaya.paudel@cea.fr}

\begin{abstract}
The final stages of mass assembly of present-day massive galaxies are expected to occur through the accretion of multiple satellites. Cosmological simulations thus predict a high frequency of stellar streams resulting from this mass accretion around the massive galaxies in the Local Volume. Such tidal streams are difficult to observe, especially in dense cluster environments, where they are readily destroyed. We present an investigation into the origins of a series of interlaced narrow filamentary stellar structures, loops and plumes in the vicinity of the Virgo Cluster, edge-on spiral galaxy, NGC~4216 that were previously identified by the Blackbird Telescope. Using the  deeper, higher-resolution and precisely calibrated optical CFHT/MegaCam images obtained as part of the Next Generation Virgo Cluster Survey (NGVS), we confirm the previously identified features and identify a few additional structures. The NGVS data allowed us to make a physical study of these low-surface brightness features and investigate their origin. The likely progenitors of the structures were identified as either already catalogued VCC dwarfs or newly discovered satellites caught in the act of being destroyed. They have the same $g-i$ color index and likely contain similar stellar populations. The alignment of three dwarfs along an apparently single stream is intriguing, and we cannot totally exclude that these are second-generation dwarf galaxies being born inside the filament from the debris of an original dwarf. The observed complex structures, including in particular a stream apparently emanating from a satellite of a satellite, point to a high rate of ongoing dwarf destruction/accretion in the region of the Virgo Cluster where NGC~4216 is located. We discuss the age of the interactions and whether they occurred in a group that is just falling into the cluster and shows signs of so-called "pre-processing"  before it gets affected by the cluster environment, or in a  group which already ventured towards the central regions of Virgo Cluster. 
In any case, compared to the other spiral galaxies in the Virgo Cluster, but also to those located in lower density environments, NGC~4216 seems to suffer an unusually high heavy bombardment. Further studies will be needed to determine whether, given the surface brightness limit of our survey, about 29 mag arcsec$^{-2}$, the number of observed streams around that galaxy is as predicted by cosmological simulations  or conversely,  whether the possible lack of similar structures in other galaxies poses a challenge to the merger-based model of galaxy mass assembly.

\end{abstract}


\section{Introduction}
Dwarf galaxies play a key role in $\Lambda$CDM-large scale structure formation scenarios, e.g. \cite{Dekel86, Navarro96}. Indeed, according to this model, the dominant process of galactic growth is the accretion of small galaxies, and the large stellar halos found around massive galaxies such as our own Milky Way have been assembled thanks to the infall (and merging with) of many small dwarf galaxies \citep{Bullock05}. Prior to their capture, dwarf galaxies are often disrupted through large tidal forces generated by their host galaxies, which form stellar streams and shells around their hosts \citep[See review by ][and references therein]{Duc12}. Although such fine structures have a lifetime of a few Gyr in isolated environments, they are much shorter lived in dense cluster environments, where they interact with the cluster potential and are quickly dispersed out  \citep{Mihos04,Tal09,Adams12}. Disrupted satellites may also be contributing stars to the intra-cluster light, where the debris generated by numerous interactions over the lifetime of the cluster is observed as a vast diffuse halo around the most massive galaxies \citep{Rudick09,Mihos05,Janowiecki10}. On the other hand, the cores of tidally stripped galaxies of intermediate mass may become compact ellipticals (cEs), and the nuclei of tidally stripped dwarfs may become ultra-compact dwarf galaxies  \citep{Sasaki07,Forbes03,Huxor11}. The scenario is commonly known as tidal threshing \citep{Bekki03}.

Because of their being stretched out, disrupted dwarf galaxies experience a strong dimming of their surface brightness and become very difficult to detect. Until recently, very few candidate ``disrupted dwarfs" were known outside of the Local Group. With the advent of sophisticated instruments and the development of data analysis techniques optimized to detect low surface brightness (LSB) structures, the number of detections is growing \citep{Forbes03, Delgado09, Delgado10, Duc11, Miskolczi11, Koch12}. However, most of the observed features -- filaments, warps, plumes and shells -- are left-overs from rather old collisions, as their progenitors are no longer visible or identifiable. The deep imaging survey of nearby spirals carried out by \cite{Delgado10} with robotic facilities such as the 0.5m Blackbird telescope show a number of cases of dwarf galaxies currently undergoing tidal disruption. These include a pair of dwarf satellites of the Virgo spiral galaxy NGC~4216 which are associated with extended tidal tails, whose brightest parts, cataloged as independent dwarf galaxies in Virgo Cluster Cataloge (VCC; \citealt{Binggeli85}), are visible on shallower SDSS images \citep{Miskolczi11}.

In this paper we revisit this system, using images obtained as part of the Next Generation Virgo Cluster Survey (NGVS, \citealt{Ferrarese12}). The excellent image quality of the survey allowed us to disclose new objects, especially a couple of filaments and disrupted dwarfs, and to perform a surface photometry and color analysis.

NGVS observations and data reduction are presented in Sect.~\ref{observations}. The results, in particular the photometric properties of the newly discovered features, are detailed in Sect.~\ref{results}. In Sect.~\ref{discussion}, we investigate the origin of the streams, their age and fate. From a global view provided by the NGVS, we consider the overall uniqueness of the NGC~4216 system within the Virgo Cluster.

\section{Observations and data reduction}
\label{observations}
The observations were carried out as part of the NGVS deep imaging survey using the MegaCam instrument \citep{Boulade03} on the 3.6m Canada-France-Hawaii Telescope (CFHT). The observational strategy, data reduction procedure and data quality control are described in \cite{Ferrarese12}. In brief, the multi-band NGVS survey benefits from the very wide field of view of the camera ($\sim$ 1$^{\circ}$ $\times$ 1$^{\circ}$), and its very good image quality (0.6"-0.9") obtained through optimizing the observing conditions. The images are processed with a dedicated data reduction pipeline, Elixir-LSB, which is explicitly designed to optimize the detection of very low surface brightness structures such as intra-cluster light (ICL), faint stellar streams and ultra-faint dwarf galaxies (\citealt{Magnier04}; Cuillandre et al., in preparation). The output products of Elixir-LSB are single frames and stacks in which the sky background has been modeled and removed (\citealt{Gwyn08}; Gwyn et al. in preparation). Further image analysis has been done using standard routines of IRAF\footnote{Image Reduction \& Analysis Facility Software distributed by National Optical Astronomy Observatories, which are operated by the Association of Universities for Research in Astronomy, Inc., under co-operative agreement with the National Science Foundation} and IDL.
 
For the study, we made use of NGVS images obtained in  the $g'$, $i'$ and $z'$ bands\footnote{noted as $g$, $i$, $z$ in the rest of the paper}. Total exposure times were 0.88, 0.57 and 1.2 hours respectively. Since the $g$-band image is the most sensitive and the cleanest, i.e., it contains the least number of artifacts such as CCD imprints and reflection halos (see below), it was preferentially used for our photometric analysis.

Surface photometry of very low surface brightness objects is notoriously difficult. The precision of their measured parameters is affected by several factors, including the accuracy of the photometric calibration, sky background variation and artifacts due to imperfect optics. The photometric calibration of the NGVS has been done using the SDSS photometry with a level of accuracy corresponding to a 1\% variation in the photometric zeropoint (see \citealt{Ferrarese12}). At the surface brightness limit of the survey (29 mag arcsec $^{-2}$ in the $g$-band), the main limitations to the LSB structure analysis and photometry are due to the reflection halos around the bright sources (stars but also the nuclei of bright galaxies with steeply rising surface brightness profiles), which show up as large disks that cannot easily be removed from the images (see an example of one of these halos in the upper-left corner of Figure \ref{main}). We therefore exclude in our analysis the contaminated areas where reflected light becomes dominant over the actual LSB feature.

Foreground stars and compact background galaxies were removed applying a ring filter to the images, and residuals were manually subtracted with the IRAF task $imedit$. The apertures used to carry out the photometry of the faint filaments and plumes, shown in Figure~\ref{label}, were defined visually as their surface brightness is too low for an automatic detection. The errors in the photometric measurements are to a large extent determined by those in the sky determination. Although a master sky background is subtracted from each image, for our aperture photometry measurements we sampled the  sky background at multiple positions around the targets. The number of sky probes ranged between 5 and 100, depending on the size of the objects. Each individual sky region consisted of  20 $\times$ 20 pixel boxes. The median value and errors were then computed.

\section{Results}
\label{results}

\subsection{NGC~4216 and  its environment}

\label{environment}

 \begin{table} 
 \caption{Basic properties of NGC~4216 and its  companion galaxies.}
 \begin{tabular}{lcccccl}
\hline
Galaxy & RA & Dec     &    M$_{B}$ & $V$ & d & Morph.\\
             & $^{o}$ & $^{o}$    & mag  & kms$^{-1}$ & kpc & \\
\hline
NGC 4216 &183.976 & 13.149&   $-$20.2       & 131 & --   & Sb(s)\\ 
NGC 4222 &184.093 & 13.307&   $-$17.1       & 230 & 56.5 & Scd\\ 
VCC 165  &183.972 & 13.215&   $-$15.6        & 255 &  19.2 & S01\\
VCC 200  &184.140 & 13.031&   $-$15.8        &  16 & 58.0 & dE2,N \\
VCC 197  &184.136 & 13.162&   $-$11.5        &  -- & 46.0 & dE1\\
VCC 185  &184.083 & 13.137&   $-$10.9        &  -- & 31.0 & dE1 \\
\hline
\end{tabular}
\tablecomments{The coordinates, heliocentric radial velocities, $V$, and morphologies are from NED. 
The absolute $B$-band magnitudes are derived from the B-band magnitudes listed in the Goldmine database\footnote{http://goldmine.mib.infn.it}. An average distance for the Virgo Cluster of 16.5 Mpc (corresponding to 80 pc arcsec$^{-1}$) was assumed,  
and d is the projected distance on the sky from the center of NGC~4216. }
\label{stb}
\end{table}

\begin{figure*} 
\includegraphics[width=18cm]{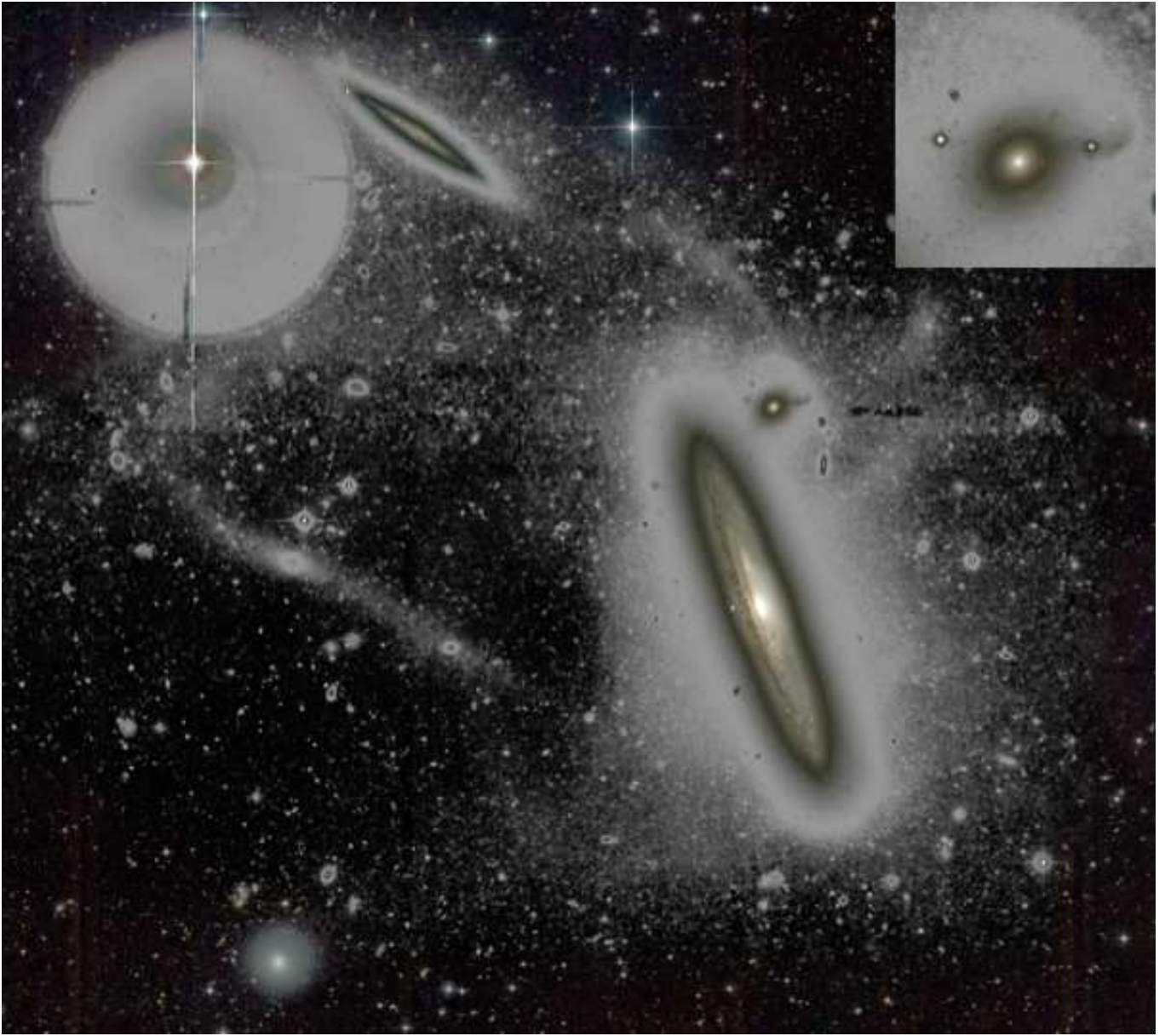}
\caption{Composite NGVS image of the field around NGC~4216. A monochromatic $g$--band image for which the faintest point-like stars have been subtracted is shown with a grey scale. For regions above a surface brightness level$~\sim$24 mag arcsec$^{-2}$, and in empty sky regions far from the main galaxy, a true color image (composite of $g$, $i$, $z$-band images) is superimposed. North is up, East is left and the field of view is 20 $\times$ 17 arcmin. The image in the inset on the upper right corner is a zoom towards VCC~165, the closest  companion of NGC~4216. }
\label{main}
\end{figure*}

Figure~\ref{main} shows a composite MegaCam image of the field around NGC~4216 (cf. Fig. 1c in \citet{Delgado10}). The regions with the highest surface brightness are displayed in ``true color" (combination of $g$, $i$ and $z$ band images) while for the low surface brightness regions, a monochromatic $g$--band image is shown. 

NGC~4216 is a barred spiral galaxy -- it is classified as SAB(s)b in the RC3 cataloge  \citep{Vaucouleurs91} --  located in the outskirts of the Virgo Cluster at an angular distance of $\sim$4 degrees from the cluster center, M87\footnote{The virial radius of Virgo's A subcluster, centred on M87, is 5.4$^{\circ}$ \citep{McLaughlin99}.}. It has an edge-on orientation with an inclination angle $i$ = 85$^{\circ}$. As shown in Figure~\ref{main}, its central regions are characterized by a prominent dust lane \citep{Chung09} and a red and compact bright nucleus. The galaxy is surrounded by a very diffuse stellar halo, and complex filamentary structures, including streams and plumes. The investigation of their nature is the focus of this study. 

The \HI\ map of NGC~4216 from the VLA \citep{Chung09} indicates an \HI\ extent which is somewhat smaller than in the optical: the ratio of the \HI\ and optical B-band radii, measured respectively at the 1 M$_{\sun}$ pc$^{-2}$ \HI\ surface density level and the 25 \masq\ isophotal level in the B-band, is $\frac{D^{iso}_{HI}}{D_{B}}$ = 0.76.

The low radial velocity of NGC~4216 (131 kms$^{-1}$, which is significantly offset relative to the $\sim$1300 kms$^{-1}$ mean velocity of the cluster itself) indicates that it is moving relatively fast with respect to the cluster center. It is the most massive galaxy within a circle of 400 kpc projected radius; the nearest galaxy of higher mass is M~99 at a projected distance of 420~kpc. Within that field of view, there are 2 early-type galaxies, and 5 late-type star-forming galaxies with radial velocity differences less than $\pm$300 kms$^{-1}$ with respect to that of NGC~4216. The basic properties of NGC 4216 and its closest companions are listed in Table~\ref{stb}. In the close vicinity of NGC~4216, towards the North lies VCC~165, at a projected distance of 19.2 kpc. VCC~165 is classified by \cite{Binggeli85} as an S0$_1$ galaxy. Another companion is located further (56.5 kpc) to the North East, at the same radial velocity as VCC~165: the spiral galaxy NGC~4222. On our deep images, the stellar halo of the galaxy NGC~4222 overlaps with the reflection halo of a bright star. South-East of NGC~4216, at 58.0 kpc distance from its centre, at a similar radial velocity, lies a nucleated early type dwarf galaxy VCC~200 \citep{Cote06}. All these companions are identified on Figure~\ref{label}.

\subsection{The streams around NGC~4216}

\begin{figure*} 
\includegraphics[width=18cm]{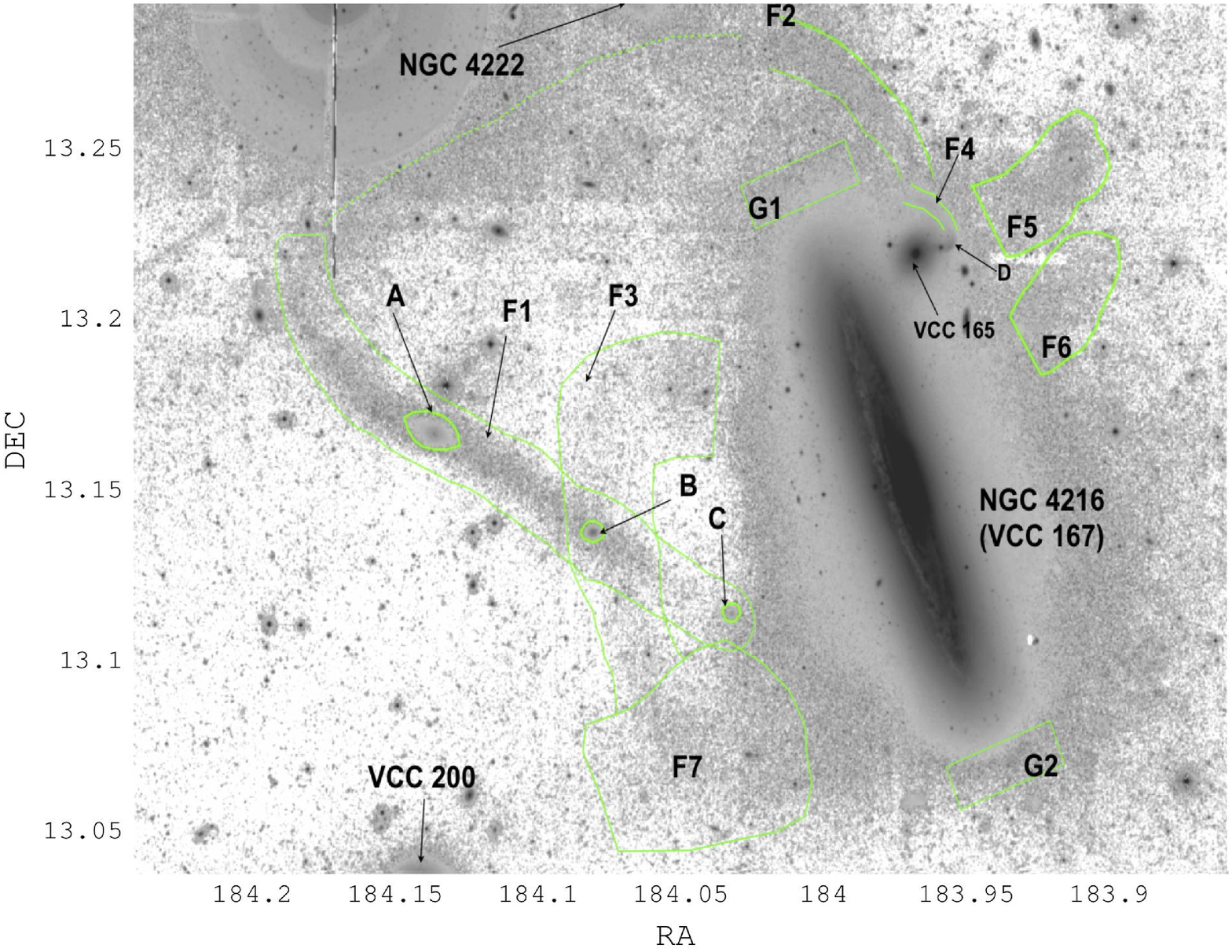}
\caption{Identification of the principal companion galaxies, stellar streams, plumes  and  progenitor candidates around NGC~4216. The apertures used to carry out the photometry are delineated on the displayed  $g$--band image. Note that for objects A, B and C, the  boundaries correspond to a surface brightness  of 27.3 mag arcsec$^{-2}$, the average surface brightness of filament F1. G1 and G2 are two selected regions in the extended stellar halo of NGC~4216. The dashed line near the extended stellar halo represents a possible, but not yet proven, connection between  F1 and F2. }
\label{label}
\end{figure*}

Figure~\ref{label} identifies the main streams and remnants of dwarf galaxies in the vicinity of NGC~4216 and VCC~165. The NGVS images clearly show the main long filament F1, which was first detected by \cite{Delgado10}. We also see the fainter structures present on the deep optical  image  of \cite{Delgado10} such as the bend in F1 towards its Easter tip, the long narrow filament to the North of the system (F2) and the plume-like filaments that stick out East of NGC~4216 (F5 and F6). Besides these previously known structures, new features are detected, such as a filament apparently wrapping around VCC~165 (F4), a loop East of NGC~4216 (F3) and a possible bridge between NGC~4216 and VCC~200 (F7). 

With respect to previous surveys, the gain in spatial resolution provided by CFHT/ MegaCam allows us to identify the dwarf galaxies presumably at the origin of these filaments. Finally, the photometric calibration precision of the multi-band NGVS  allowed us to determine the photometric properties and colors of these various features, which so far had not been possible. The photometric  data on the streams and progenitors are presented respectively in  Table \ref{ftb} \& \ref{ctb}. \\

\begin{figure*} 
\includegraphics[width=18cm]{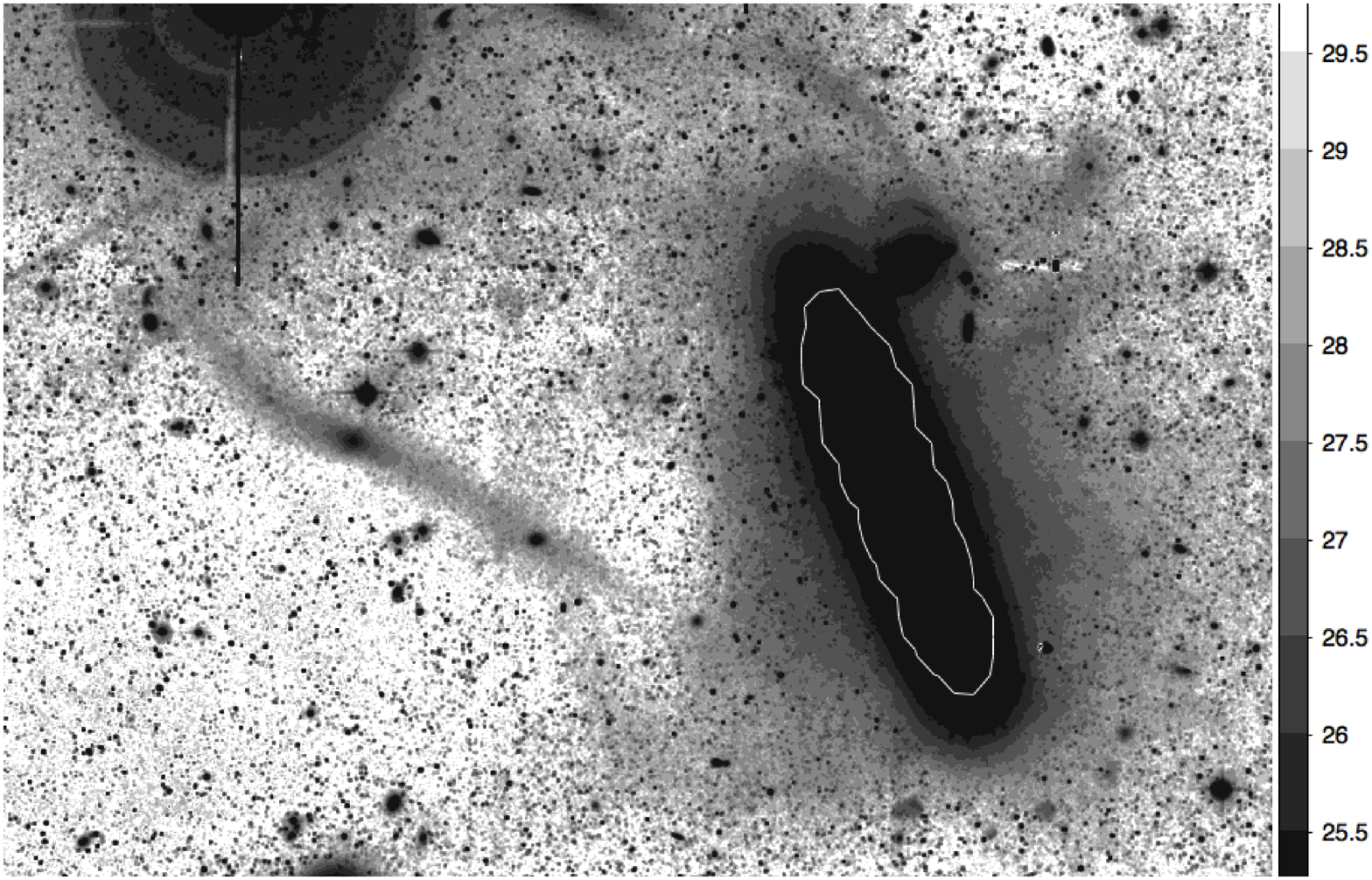}
\caption{$g$-band surface brightness map of the field around NGC~4216. The surface brightness scale in mag arcsec$^{-2}$ is shown to the right and the field of view is the same as for Figure~\ref{label}. The white contour corresponds to the faintest level of 21-cm \HI\ line  emission in NGC~4216 detected with the  VLA  by \cite{Chung09}.} 
\label{sfb}
\end{figure*}

\begin{figure*}
\includegraphics[width=16cm]{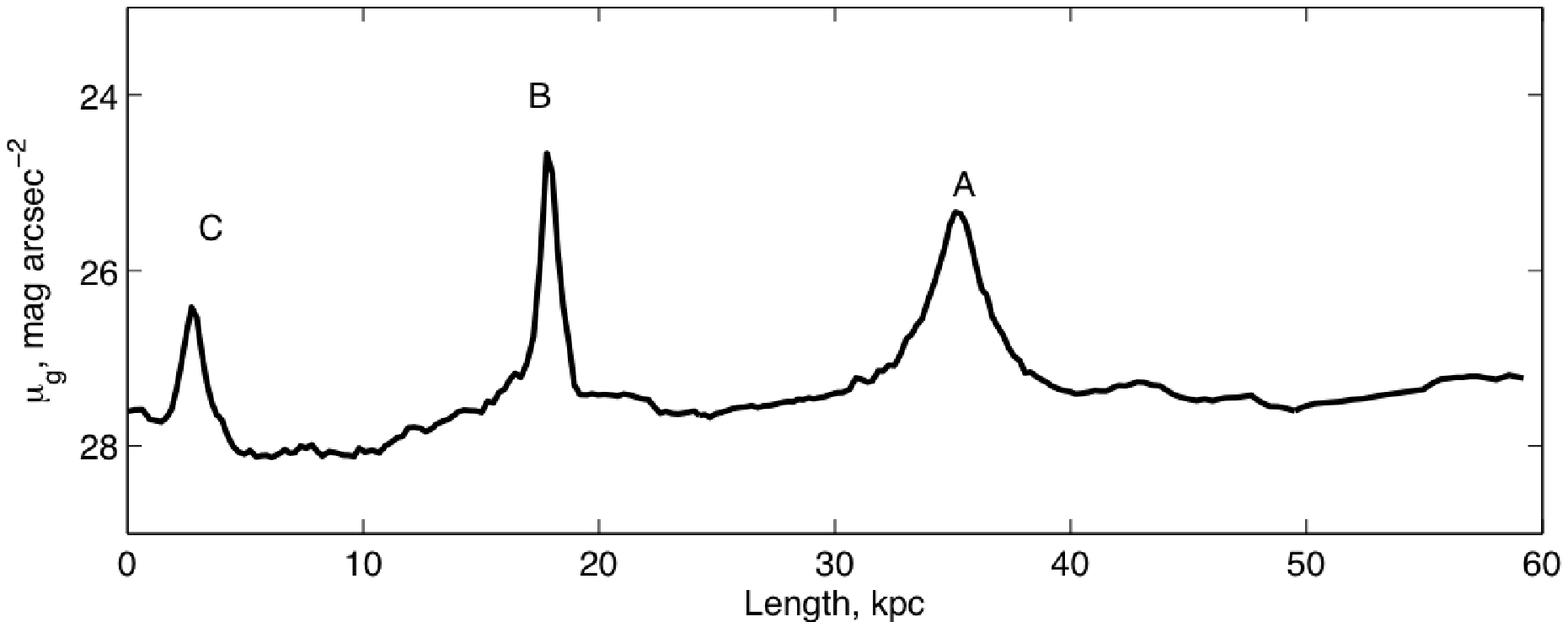}
\caption{Variation of the median $g$--band surface brightness along the length of filament F1,  from dwarf galaxy C near NGC~4216 to  its easternmost tip where it gets lost within the halo of star. Photometric measurements were obtained collapsing the filament along its width and median-combining the data. The positions of the dwarf galaxies  A, B and C (see Fig. 2) are indicated.}
\label{sfbp}
\end{figure*}

The most prominent tidal stream, F1, East of NGC~4216, has an angular extent of $\sim$12.5 arcmin on the sky, which corresponds to $\sim$60 kpc at the assumed mean distance of the cluster, 16.5 Mpc \citep{Mei07}. The filament starts from the South-East corner of NGC~4216 and near its easternmost tip it makes a sudden turn towards the West-North with an angle of about 120 degree. Although further away it gets lost within the halo of a bright red star, when it emerges again, it seems to join the filament F2, the second longest filament in the field, North of NGC~4216. Unfortunately, the stellar halo glare in the region between the two filaments does not allow us to say with certainty that a physical connection exists. No such bridge is visible on slightly shallower image of \cite{Delgado10}, where the stellar halo is less obtrusive. The width of the stellar stream F1 appears constant along most of its length, about $\sim$3 kpc. It broadens after the bend. As shown in Figure~\ref{sfbp}, the average surface brightness of the filament (excluding regions close to the dwarf galaxies along it) is slightly below 28 mag arcsec$^{-2}$. 

The NGVS image discloses another rather broad stream, F3, with a surface brightness below 28.5 mag arcsec$^{-2}$ (See Figure~\ref{sfb}), which sticks out of NGC~4216, then bends to cross filament F1 and finally falls back towards the host galaxy, near  F7. 

The very broad filament F7 points in the direction of the dwarf companion VCC~200, but a physical connection between the two galaxies cannot be firmly established. The plume-like stellar structures which apparently stick out from the main body of NGC~4216 on its western side (F5 and F6) may in fact be filaments which connect  to  F1--F3   behind the galaxy. They   have a nearly uniform surface brightness of 27.5 mag arcsec$^{-2}$.

Other filamentary structures can be seen (best on Figure~\ref{dwarf}) around VCC 165, the closest companion of NGC 4216.
On the displayed g-band image, a model of VCC~165 obtained with the IRAF task {\it bmodel} has been subtracted to further enhance the filaments. The most prominent one,  the 6-kpc long structure labeled F4, wraps around the dwarf companion on one side, and points towards the galaxy center on the other. There is a further hint that the galaxy extends further towards the East in the direction of the spiral.  VCC~165 itself, despite its apparent proximity to NGC~4216, does not exhibit strong signs of morphological perturbation. The map obtained subtracting the ellipse model (see Figure~\ref{dwarf}) does not show prominent residuals in its outer regions. In the inner regions, spiral-like features (or narrow dust lanes) may be observed. \\

We measured the $g-i$ colors for the most prominent features. The $i$--band photometric measurements were particularly delicate as this band shows several residuals at low  surface brightness levels, such as chip gaps. The contamination by stellar halos is also more severe in this band. Even the nucleus of NGC~4216 generates an artificial, disk--like structure with a radius of nearly 3~arcmin, which is almost invisible in the $g$--band, but becomes noticeable in the $i$ and $z$ bands. Regions affected by the stellar or galactic reflection halos were excluded from our analysis. The colors are listed in Table \ref{ftb}. The  $g-i$ color  for F1 is uniform along its  length and has an average value $g-i$ = 0.9 $\pm0.1$ mag. Interestingly, this value agrees, within the errors, with that of filament F2, possibly suggesting a common physical origin for the two filaments. On the other hand, there is a significant difference of 0.4 mag between filaments F4 and F2, although both structures touch each other on the projected image. As argued later, they may have different progenitors, and indeed their different orientation supports this hypothesis.

\begin{figure} 
\includegraphics[width=9cm]{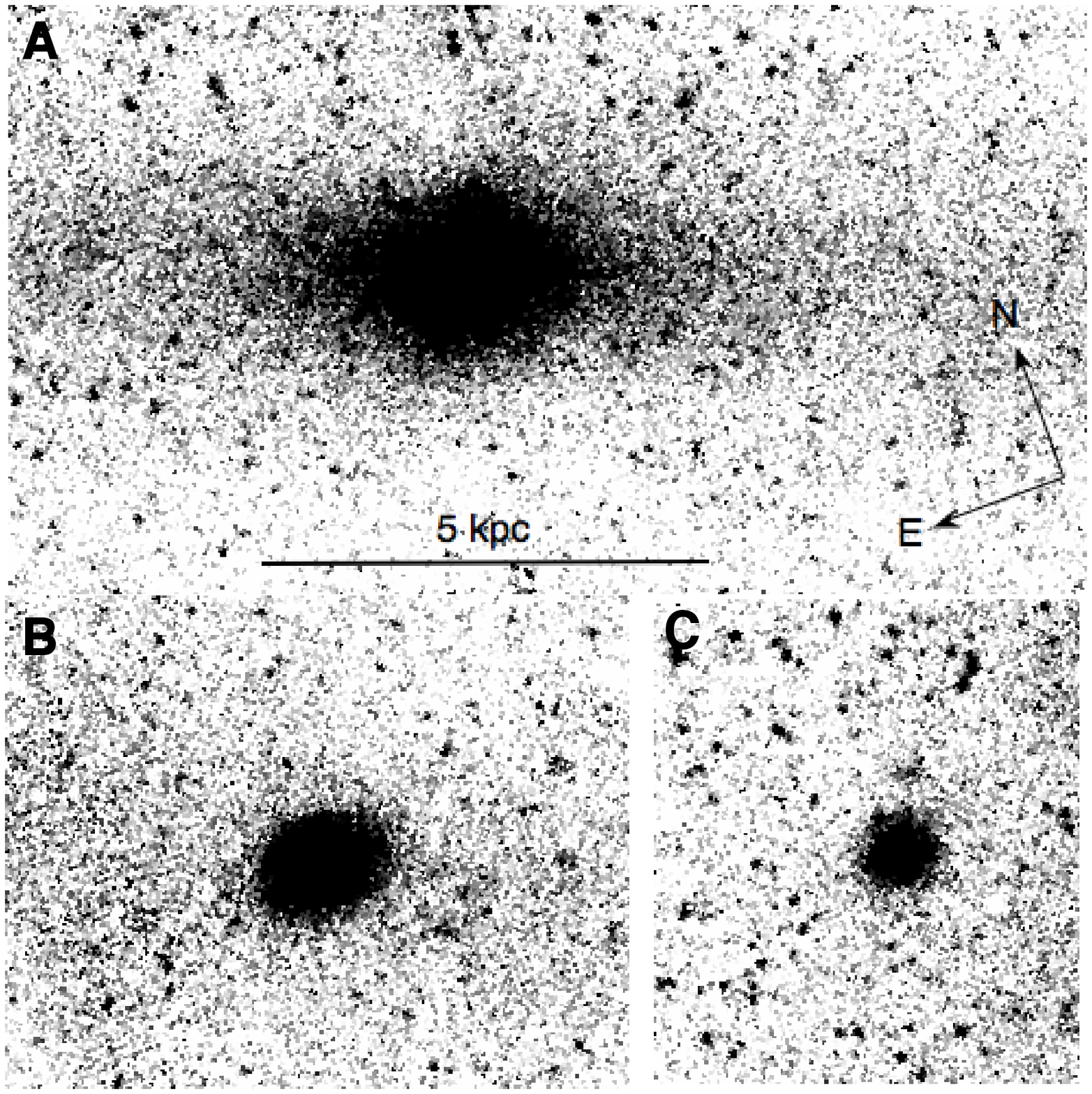}
\includegraphics[width=9cm]{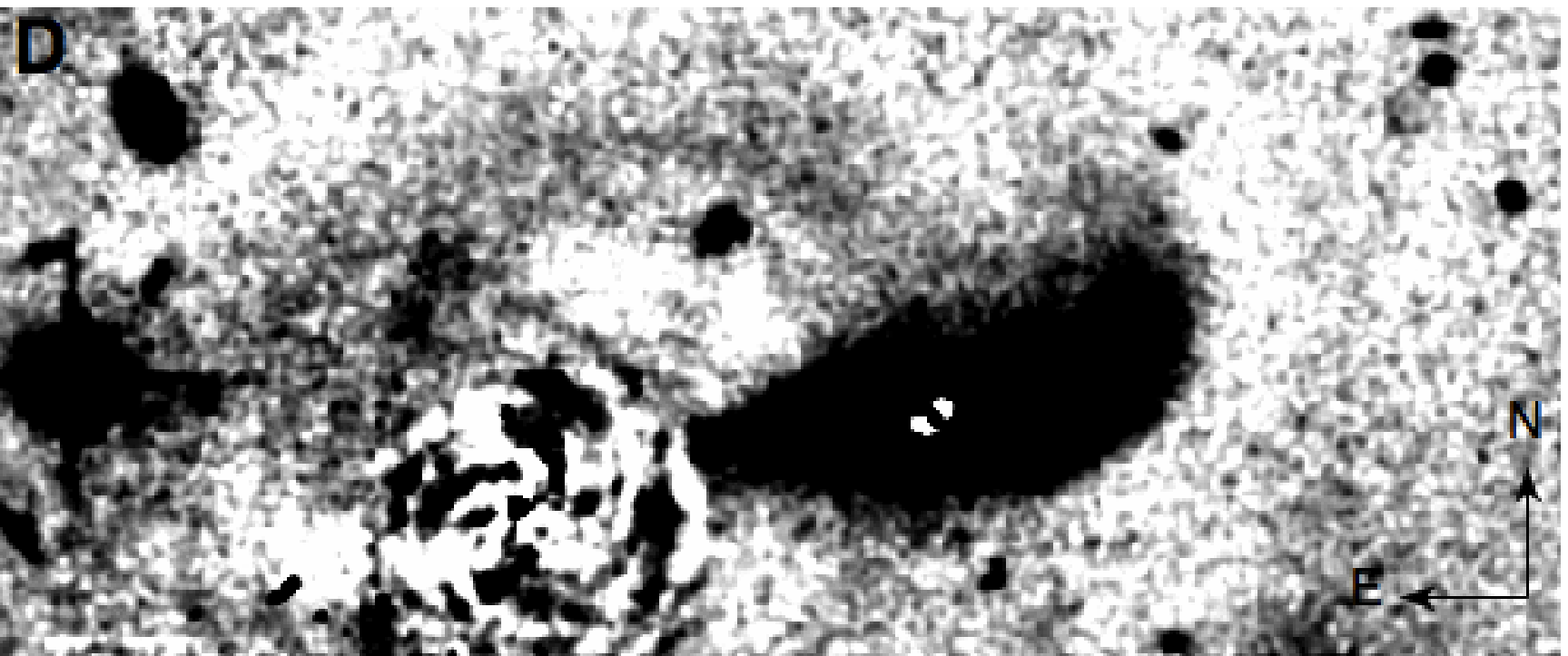}
\caption{$g$--band images of the disrupted dwarfs A, B, C  and D.   For objects A, B and C, located along filament F1, the  images have been rotated  such  that the local orientation of the filament is parallel to the horizontal axis. 
D is a disrupted dwarf galaxy candidate located West of  VCC165. On the displayed image, the latter galaxy has been subtracted using an ellipse model. All images are displayed with the same intensity and spatial scales.}
\label{cores}
\end{figure}

\begin{table}
\centering
\tabletypesize{\scriptsize}
\caption{Properties of the filaments (F1-F7) and external halo regions (G1-G2) of NGC~4216.}
\begin{tabular}{c c ccl}
\hline

Feature & $g-i$ & $< \mu_{g} >$ & $M_{B}$ & log($M_{*}$)\\
& mag & mag.arcsec$^{-2}$  & mag&M$_{\sun}$\\
\hline
F1 & 0.9 $\pm$ 0.1& 27.8 $\pm$0.1& -14.2& 8.2\\
F2 & 0.7 $\pm$ 0.1& 26.7 $\pm$0.1& -12.2& 7.1\\
F3 & --                    & 28.6 $\pm$0.5          & -13.5& 7.8\\
F4 & 1.1 $\pm$0.1 & 25.8 $\pm$0.1& -10.0& 6.9\\
F5 & 0.7 $\pm$0.2 & 27.1 $\pm$0.1& -13.0& 7.5\\
F6 & 0.8 $\pm$0.1 & 26.9 $\pm$0.1& -13.0& 7.6\\
F7 & 1.0 $\pm$0.4 & 27.4 $\pm$0.2& -14.2& 8.2\\ \hline
G1 & 0.8 $\pm$0.1 & 26.0 $\pm$0.1&       &     \\
G2 & 0.7 $\pm$0.1 & 26.4 $\pm$0.1&       &     \\

\hline
\end{tabular}
\tablecomments{The color $g-i$ was determined from aperture photometric measurements in areas not contaminated by instrumental artifacts. $<\mu_{g}>$ is the average surface brightness within the areas defined in Figure~\ref{label}, excluding the progenitors  (i.e. A, B \&  C).
 The associated errors are computed from  the standard deviation of the individual sky background measures.
 The absolute blue magnitude in the $B$ band,  $M_{B}$,  was derived from the extrapolated observed magnitude, i.e., the average surface brightness times the area, converting the MegaCam magnitudes to the  standard Johnson system using the calibration of \cite{Jordi06} and assuming a distance of 16.5 Mpc \citep{Mei07}. The stellar masses,  $M_{*}$, were estimated from the extrapolated absolute magnitudes assuming the mass to light ratios of \cite{Bell03}.} 
\label{ftb}
\end{table}

\subsection{Stellar cores within the  filaments}
Several objects of higher surface brightness are visible towards the streams: such cores are likely the remnants of tidally disrupted dwarf galaxies and as such appear as  good progenitor candidates of the filaments' stellar material. 

As shown in Figures \ref{label} and \ref{sfb},  the most extended filament, F1, hosts three relatively compact stellar objects.
 The most luminous object, A, is also the most extended (see Figure~\ref{cores}). It has a $\sim$24.5 mag arcsec$^{-2}$ central $g$-band surface brightness, which in principle is high enough to have it included in the Virgo Cluster Catalog (VCC~197). Modeling the galaxy with ellipses (using the IRAF task $ellipse$ and $bmodel$), we found that it is stretched out along filament F1 and contains a bar-like structure along the minor axis, and a compact nucleus. Modeling  the galaxy with a  S\'ersic model with GALFIT \citep{Peng02}, we obtained a best fit with an exponential profile and a very large effective radius of $\sim$5 kpc, a consequence of the tidal threshing. 

Object B -- also in the VCC (VCC~185) -- with  an effective radius of 0.4 kpc and an ellipticity of 0.29, is more compact and rounder. Although located at the crossing of filaments F1 and F3, its major axis is aligned with neither of these streams. Finally, object C is an uncataloged faint dwarf located near the base of the tail. It has a round shape and does not show any clear evidence of a tidal interaction. The integrated  $g-i$ colors of  both A and B are in very good agreement with that of filament F1, while C is 0.2 mag bluer. Given their rather red colors, none of the galaxies show evidence of containing young stars. A is not detected on a deep 2600 sec archival GALEX/NUV image \citep{Boselli11}, while B shows very weak UV emission. At Arecibo, the ALFALFA \HI\  survey \citep{Haynes11} did not detect any atomic hydrogen emission towards the three passive dwarfs. The much deeper AGES fields did not cover the NGC 4216 area \citep{Taylor12a, Taylor12b}.

\begin{figure} 
 \includegraphics[width=8cm]{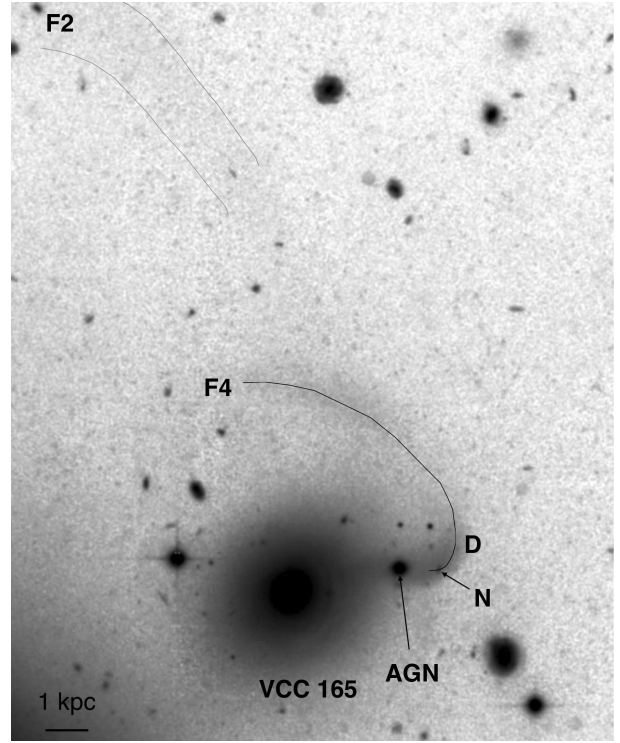}
\caption{$g$--band image of  the field around  VCC~165.  The main features have been labeled: the streams F2 and F4, the progenitor candidate of F4, the disrupted dwarf D, with its possible nucleus, N, and  the background AGN. See the inset in Figure \ref{main} for another view in color on this complex structure.}
\label{dwarf}
\end{figure}

\begin{figure} 
\includegraphics[width=8cm]{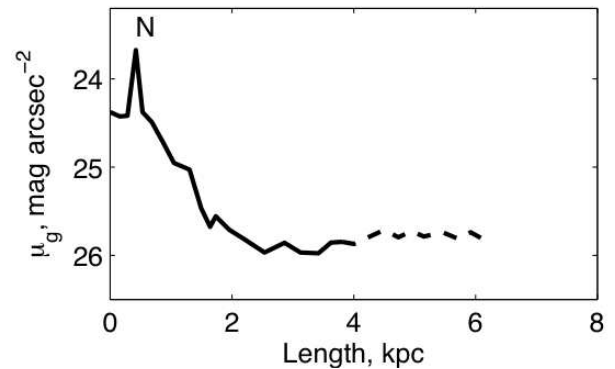}
\caption{Surface brightness profile along filament F4. The black solid curved line in Fig \ref{dwarf} delineates the region traced for the profile.  }
\label{dprof}
\end{figure}

The brightest condensation along Filament F4, West of VCC~165, turns out to be a chance superposition of a background low--mass AGN at a redshift of 0.043, according to its SDSS spectrum \footnote{The AGN, SDSS J121551.26+131303.4, belongs to a rich large-scale filament of galaxies. Its emission lines classify it as a Seyfert 2. With an absolute $g$--band magnitude of $-$18.2, it is among the least massive AGNs found in the whole SDSS survey. A visual inspection of all spectroscopically confirmed SDSS AGNs below  z $<$ 0.05, less luminous than  Mg $>$ -19 mag, did not reveal  other objects with  tidal tails apparently associated to them.}. The second brightest object in that region is an extended structure that lies in the continuation of F4. Given its much higher surface brightness than that of filament (see  Figures~\ref{cores} and \ref{dprof}), we argue here that it could be in fact the main-body remnant of an almost completely disrupted dwarf, which we have labeled on the figures as D. The fact that D and F4 have similar colors is consistent with this hypothesis. A small compact central core, labeled as N, can be seen towards D; it has the  same color and could be the nucleus of the latter. The $B$--band absolute magnitude of D -- within the boundaries defined in Figure \ref{dwarf} -- is -11.3  mag, i.e. in between that of dwarfs A and B.

The  mass of the stars still bound to the dwarfs (A+B+C+D) amounts to less than one tenth of the total  stellar mass of the filaments (F1-7).  Assuming that all the stellar material in filament F1 was pulled out from galaxy A,  then the  galaxy must originally have had an absolute magnitude of M$_B$ = $-$15.6 mag, a value in the range of bright early-type dwarf galaxies in the Virgo Cluster \citep{Lisker07}.

\begin{table} 
\centering
\tabletypesize{\scriptsize}
\caption{Properties of the  dwarf galaxies identified along the filaments. }
 \begin{tabular}{cccccc}
\hline
Object  & $M_{B}$ & $g-i$ & e & log(M$_*$) & R$_e$\\
        & mag     & mag   & --  & M$_{\sun}$ & kpc \\
\hline
A(VCC 197) & $-$12.1 & 0.88 $\pm$ 0.02 &0.79 &7.2 & 4.5\\
B(VCC 185) & $-$10.2 & 0.86 $\pm$ 0.02 &0.29 &6.7 & 0.4\\
C          & $-$09.6 & 0.62 $\pm$ 0.04 & --  &6.0 & 0.3\\
D 	       & $-$11.3 & 1.05 $\pm$ 0.02 & --  &7.1 &  --\\
(N) 		   & $-$07.0 & 1.05 $\pm$ 0.04 &0.63 &5.3 &  --\\
VCC 165    & $-$15.5 & 1.22 $\pm$ 0.01 &0.15 &8.9 & 1.3\\
\hline
\end{tabular}
\tablecomments{ 
The absolute blue magnitude and colors were determined from  photometric measurements made within the apertures defined in Figure~\ref{label}. The ellipticities $e$  were derived using the IRAF $ellipse$ task. The effective radius are derived with a 2D fitting of  the light profile with  a S\'ersic model made with GALFIT \citep{Peng02}. D is the disrupted dwarf galaxy near VCC 165 (see Fig. \ref{dwarf}), while N is its candidate nucleus. }
\label{ctb}
\end{table}

\section{Discussion}
\label{discussion}

In the Carnegie Atlas of Galaxies \citep{Sandage94}, the spiral galaxy NGC~4216 is described as ``one of the most famous galaxies in the sky because it is often used in textbooks to illustrate the bulge, disk, dust content, and spiral pattern of typical luminous galaxies near the middle of the Sb classification sequence". The deep NGVS images we have presented here show many additional features around this prototypical galaxy, in particular a complex network of interlaced plumes, streams and filaments. Although a detailed interpretation is not straightforward, the large number of low surface brightness stellar tails is consistent with predictions from numerical simulations about the mass assembly of galaxies \citep{Johnston01, Bullock05}. This raises the question why all massive galaxies do not show a similar level of fine-structures. In the following, we propose scenarios accounting for the various features seen around NGC~4216 and discuss what is so special (or not) about this galaxy.

\subsection{Real structures or artifacts}
Disentangling real features from artifacts is essential to the goals of this paper. Although multiple internal reflections in CCDs and stellar halos also produce low surface brightness structures, these instrumental features usually have somewhat regular geometric shapes, like disks, long thin straight lines or grids, while the shapes of the cosmic structures of interest to us  are significantly different in appearance: broad filaments, arcs and irregular plumes. In the case of NGC 4216, a number of the faint structures we noted are (barely) visible on SDSS images \citep{Miskolczi11} and more convincingly detected by \cite{Delgado10}. 

As a reminder, the principal features observed around NGC~4216 and possibly associated with collisional debris are the following:\\
(a) two long filaments (F1 and F2), with frail evidence that they are physically connected, forming a possible loop in the North-West quadrant of NGC 4216. Three dwarf galaxies (A, B and C) are found projected onto, and we argue, associated with filament F1; \\
 (b) a less extended very low surface brightness arc-like structure east of NGC4216 (F3). The dwarf satellite B is found at the intersection of F3 and F1; \\
 (c) a relatively high surface brightness filament wrapping around the dwarf galaxy companion VCC~165, likely emanating from a newly discovered disrupted dwarf (D); \\
 (d) several broader and short filaments coming out from main body of NGC~4216 (F5-7).

\subsection{Origin of the filamentary structures}
Undoubtedly, all the stellar filamentary structures observed in the field of NGC~4216 have been shaped by tidal forces. However the origin of their stars may be debated. Given the high frequency of fine structures around the spiral, one may wonder if they were all produced by a single event: either an old merger with a relatively high mass companion, or ongoing tidal interaction with dwarf galaxies.  In these two cases, the dwarfs seen along filament F1 would not be the original progenitors of its stars, but they would instead be tidal dwarf galaxies (TDGs) born in situ from the collisional debris. In fact, the three aligned dwarfs A, B and C show similarities with the three also aligned TDG candidates recently discovered along a prominent tidal tail associated with the elliptical galaxy NGC~5557, which was interpreted as an old merger \citep{Duc11}. However the latter are gas-rich, contrary to the NGC~4216 dwarfs. 

NGC~4216 has at least one relatively massive companion, the edge-on spiral NGC~4222, and two luminous dwarf satellites, VCC~165 and VCC~200 (see Table \ref{stb}). The latter two galaxies are likely not massive enough to tidally disrupt their host (though the filament F7 pointing towards VCC~200 is intriguing). An on-going collision between the two spirals would generate tidal tails in the plane of their stellar disks (see review by \citealt{Duc13}). In fact, all the new features discovered around the galaxy (the stellar protuberances, plumes and  filaments) are located along its minor axis, i.e., perpendicular to its main plane, and not towards the outermost regions of its stellar disk plane, where the isophotes remain regular (see in particular Figure~\ref{main}). In \HI, \cite{Chung09} did not find any clear indications of interactions between the two galaxies either, though they noted mild distortions in the kinematics at the edge of the disk in both systems.  A major merger in NGC~4216 is thus very unlikely. Furthermore,  the outer regions along the major axis of the galaxy, G1 and G2 (see Figure~\ref{label}), are slightly bluer in $g-i$ by 0.2~mag than the regions along the intergalactic stellar streams. This likely means that the stellar streams and outer disk of the galaxy do not share the same stellar populations, and thus that the stars in the filaments did not originate from the spiral.

Alternatively, the filamentary structures may result from dwarf galaxies tidally stirred by NGC~4216. In that case, the tails are shaped by both tidal forces and the orbit of the satellite around the host galaxy \citep{Varghese11}. Inside several of the filaments, stellar bodies with higher surface brightness are found that are good filament progenitor candidates. The most promising one is object A, which shares with filament F1 its orientation and color. Scrutinizing Figure~\ref{main}, one may see that the filament has a S-shaped structure close to the dwarf, due to the presence of two slightly misaligned tails on each side of the galaxy, possibly the leading and trailing tidal tails. This provides an additional evidence that A is the progenitor. 

On the other hand, it is hard to determine whether object B belongs to filament F1 or F3, or if it is just a chance superimposition. Its main body is not aligned with any of the filaments while its $g-i$ color agrees with that of filament F1 (the S/N of F3 is too low to determine its color). If objects B, as well as C, are indeed associated with F1, their presence along a single filament remains mysterious. It is unlikely that their common progenitor broke in two or even three stellar cores. Alternatively, if A, B and C were all pre-existing objects, and subsequently tidally disrupted by NGC~4216, one would a priori expect them to be originally on different orbits and therefore to generate filaments with different orientations. Note however that according to cosmological models, satellite accretion occurs along privileged directions. The dwarfs might thus belong to a common cosmological stream. A chance superimposition of B and C is obviously an hypothesis worth exploring. Having spectroscopic data, and in particular information on the radial velocity and chemical abundances, would help to determine the nature of the dwarfs. Alternatively, B and C might be substructures born in situ in the tidal tail of A. Such a formation mechanism would differ from the one at the origin of TDGs, as it would not involve  a major wet merger. It would rather resemble the one responsible for the formation of the stellar clumps that are commonly found in the tidal streams of Milky-Way Globular Clusters \citep[e.g.,][]{Mastrobuono-Battisti12}. Note that in our own local group, the distribution  of dwarf satellites inside narrow disks, the so-called disks of satellites,   \citep[e.g.][]{Pawlowski12,Ibata13}, is also puzzling and no satisfactory solution to account for them  has yet been found. \\

Finally, the origin of the filamentary structures that do not show any obvious progenitor, such as F2 (unless it is connected to F1 and shares the same progenitor), and of the three plumes-like filaments (F5-7) is even less constrained. The progenitors might have been totally disrupted or their remnants might already have been swallowed by the massive spiral. In fact, object D near VCC~165 is likely a case of a dwarf in the process of total disruption. Whether it is disrupted following its interaction with VCC~165 (around which its tidal tail, F4, wraps) or with the more massive but slightly more distant galaxy NGC~4216, or both, is unclear. VCC~165 itself does not seem to be affected by gravitational interactions.

\subsection{An unusual on-going bombardment in a cluster of galaxies}
 
The discovery of numerous filamentary structures around NGC~4216, and tidally stretched stellar clumps within them, revealed an impressive number  of in-falling satellites caught it the act of disruption. When has such a bombardment started and where in the cluster did it occur? 

\subsubsection{A sub-structure within the Virgo Cluster?}
Could  NGC~4216 belong to a sub-structure, like a group that has  been accreted by the Virgo Cluster? Within the cluster, NGC~4216 is located in a moderately dense environment: the local projected number density of galaxies is nearly half that of the cluster core. The spiral is the most luminous galaxy within this sub-structure. For each Virgo galaxy more luminous than M$_{B}$ =  -18, we estimated in the SDSS dataset the number of companions within a 100 kpc radius and radial velocities of $\pm$300 kms$^{-1}$. NGC 4216 turns out to have a significantly higher number of faint companions (M$_{B}$ $>$ -18, with spectra available), compared to its fellow galaxies. The probability of an increased number of dwarfs being demolished is thus higher there. The statistics on the ratio of dwarfs to giants is still to be extended to the many faint and ultra-faint dwarfs that are being discovered by the NGVS survey. Low velocity collisions able to generate multiple tidal tails are expected in groups. As a matter of fact, examples  of tidal interaction and possibly ram--pressure have been observed in several in-falling groups of clusters, which are suffering a so--called ``pre-processing" \citep[e.g.,][]{Cortese06}, during which the properties of their galaxies are changed within the group before being affected by the cluster environment.

\subsubsection{Time scales for the satellite accretion/destruction}
A time scale  for the  satellite accretion/destruction process may be obtained from the survival time of  the debris: whereas around isolated systems, streams may remain visible for several Gyrs, within a cluster,  the life expectancy of tidal debris is largely reduced due to dynamical heating and mixing \citep{Rudick09}.  They  should evaporate on a time scale shorter than the cluster crossing time, typically 1 Gyr  \citep{Trentham02}. 
This suggests that  the  bombardment  has either started within the Virgo Cluster itself or in a pre-processed group that was accreted by the cluster only recently. 

The gas content of the NGC~4216 system somehow differs from that of the pre-processed groups so far reported in the literature. The filamentary structures detected in claimed pre-processed groups \citep[e.g.,][]{Cortese06} are usually gas--rich, host \HII\ regions, and are thus probably star-forming.  There is no evidence for intergalactic \HI\ gas around NGC~4216, at least in the ALFALFA single-dish survey\footnote{Considering that some large-scale gas features of low column density such as the Virgo \HI\ plume discovered by the WSRT \citep{Oosterloo05}  have been missed in single dish surveys, deeper imaging studies with interferometers which will not resolve out diffuse gas will help to ensure the absence of intergalactic gas.} 

The tails are red, presumably made of old stars. The dwarf satellites around NGC~4216 had their gas stripped long ago, either due to long-lasting interactions with the massive galaxies in the group, or because globally the group members had their gas reservoir stripped before the accretion event. As noted in Sect.~\ref{environment}, the  \HI\  disk of NGC~4216 seems truncated, which is normally interpreted as a consequence of ram pressure stripping in the cluster environment: the outer gaseous halo of a spiral galaxy is stripped away when it passes through a high density medium such as the core of a cluster  \citep{Vollmer01,Boselli06}. This might indicate that the NGC~4216 group already suffered  a ram-pressure stripping episode, and therefore that is has already been in the cluster for a long time. The ram pressure needed to strip a spiral down to  the observed outer \HI\ radius $R_{HI}$ (15~kpc) is $\sim \Sigma_{HI} V_{rot}^2/R = \rho_{ICM} V_{gal}^2$, where $\Sigma_{HI}$ is the \HI\ surface density (3 M$_{\sun}/pc^2$), $V_{rot}$ the rotation velocity (250~km s$^{-1}$) and $V_{gal}$ the galaxy velocity in Virgo cluster (1500~km s$^{-1}$). This yields an ICM density, $\rho_{ICM}$ of $3 \times 10^{-4}$ cm$^{-3}$. According to \cite{Schindler99}, such a density occurs at a cluster radius of about 0.5 Mpc or 1.7 degrees. At the present epoch, NGC~4216 is located at projected distance of 2.5 and 3.6 degrees respectively from the central cluster galaxies, M86 and M87. Thus, the value of its \HI\ truncation radius suggests a stripping episode that occurred at least  1 Gyr ago, given the galaxy velocity, and closer to the cluster core, which would imply that the galaxy has ventured through that environment in the past\footnote{The true distance and location of NGC~4216 within the Virgo Cluster is not known. An estimate of the distance based on Surface Brightness Fluctuations is only available for one possible group member,  VCC~200. \cite{Blakeslee09} determined a distance of 18.2 $\pm$0.6 Mpc for this galaxy. The group thus may be located  further  from M87,  at $\sim$2 Mpc (calculated 3D distance using radial and projected distance 1.7 and 1.2 Mpc respectively).  This yields a travel  time  from the centre of Virgo Cluster to the current position of nearly 2 Gyr.}. The bombardment of NGC~4216 and the multiple dwarf destruction must  then have occurred later on.

Alternatively,  the \HI\ truncation of NGC~4216 might be possible without invoking the Virgo Intra Cluster Medium, if the  group were sufficiently massive and X--ray luminous  \citep{Sengupta07,Kern08}. There is however no evidence for extended X-ray emission at the location of the galaxy to support the hypothesis of an in situ ram pressure event   \citep[See Figure~7 in ][]{Chung09}.
Finally one should note that one of the main characteristics of the gas property of NGC~4216, more striking than the disk truncation, is its overall   low  \HI\ surface density  for a spiral galaxy of its size. Such a  global depletion of \HI\  might in fact be due to the effect of turbulent viscous stripping \citep{Nulsen82,Chung09} rather than ram pressure stripping.

Taking furthermore into account the fact that the  companion galaxy, NGC~4222,  has a regular \HI\ distribution,  it is still difficult to firmly conclude that the group of NGC~4216 did experience a ram pressure event during a past journey through or near the Cluster Core.

\subsubsection{How unique is NGC~4216?}
How special is NGC~4216 within the Virgo Cluster? As part of the NGVS project, we have identified in a systematic way low-surface brightness stellar tails, and cases of tidally disrupted dwarfs. Our preliminary analysis over most of the Virgo Cluster area indicates that  there are only a  rather small number of them, especially around spiral galaxies (Duc et al., 2013 in prep.). This may be an additional argument in agreement with a recent accretion of NGC~4216 as part of a group. 

On the other hand, even in the field and in groups, a bombardment like the one suffered by NGC~4216  seems pretty rare.
None of the spirals mapped by  \cite{Delgado10} have a similar number of filaments and disrupted dwarfs around them. In some systems, what appears to be multiple loops may in fact  be due the wrapping of a single tidal stream around its host  \citep[e.g.,][]{Delgado09}. In the case of NGC~4216, several independent filaments are observed; they  emanate from at least 2, likely 4, different disrupted dwarfs. The on-going deep imaging survey with MegaCam of Atlas3D field galaxies \citep{Duc11}, which has a similar depth and sensitivity to low-surface brightness structures as the NGVS, has so far also failed to detect the same wealth of debris of minor mergers observed around  NGC~4216. 
On the other hand, multiple stellar streams and  dwarf satellite progenitors have been identified in the halo of our own Milky Way \citep{Ibata01,Yanny03,Grillmair06}. They are however of much lower surface brightness. Were the NGVS able to reach the surface brightness limit achievable with resolved stellar populations -- 32 instead of 29 mag arcsec$^{-2}$  --  then the number of filaments and satellites is expected to strongly increase according to numerical simulations  \citep[][Michel-Dansac et al., 2013, in prep.]{Bullock05}.

\section{Conclusions}
We have presented a study of the properties and origins of a complex series of interlaced narrow filamentary stellar structures, loops and plumes located in the vicinity of the edge-on Virgo Cluster spiral galaxy NGC~4216. They were detected on multi-band extremely deep optical Next Virgo Cluster Survey (NGVS) images. Some streams had already been identified in previous surveys, e.g. \citet{Delgado10}.  Several others, with $g$--band surface brightness levels fainter than 28.5 mag arcsec$^{-2}$ are reported here for the first time.\\

Whereas the tidal origin of the filamentary structures seems rather obvious, determining the progenitors of their stars is less straightforward. The high resolution of the MegaCam camera, besides its high sensitivity to low surface brightness features, allowed us to identify progenitor candidates and to further explore the complexity of the  system. This is  illustrated by the detection of satellites of satellites: a filament was found wrapping itself around the luminous dwarf VCC~165, which itself is probably orbiting around NGC~4216.  The alignment of three dwarfs along an apparently single stream is also intriguing. Projection effects might be an explanation  unless   some of them are second-generation objects formed in-situ in tidal debris,  or if all the dwarfs were accreted from the same cosmological filament.

The dominant spiral itself does not show any strong evidence of having experienced a recent merger or being currently involved in a tidal interaction with a massive companion. The stellar streams most likely result from the disruption of dwarf satellites, some of which are still visible but look highly highly disrupted on the NGVS images.\\

While dwarf galaxy accretion, with an associated high rate of satellite destruction and the formation of stellar streams, is a priori expected for massive galaxies -- cosmological semi-analytical models and simulations predict that they grow at low redshift through minor mergers \citep{Naab09,Khochfar11,McLure13} - their presence in a cluster environment is rather surprising. A possible explanation is that NGC~4216 is the central galaxy of a group that has recently fallen into the Virgo Cluster, and that the merging activity seen in its surroundings might be considered as evidence for  pre-processing. This scenario  does not fully explain  the apparently truncated \HI\ distribution of the spiral galaxy, however, if caused by a ram-pressure stripping event. The level of ram-pressure required to account for the \HI\ deficiency might  imply that the spiral  has already passed closer by  the central regions of the cluster at least $\sim$1 Gyr ago. This hypothesis would raise the question of the survival of  stellar streams in the cluster environment, which might be longer than once believed. However, other processes  that do not request an interaction with the central ICM are also able to truncate the \HI\ disk. Follow-up observations, in particular a spectroscopic survey to get dynamical information on the system, would be  needed to determine how long ago the group of NGC~4216, including its numerous satellites, joined the Virgo Cluster.\\

We have carried out a preliminary comparison of the fine structure index of  NGC~4216 with that  determined in other spiral galaxies in the Virgo Cluster, but also around massive galaxies in the field and in groups. NGC~4216  seems to have a particularly high number of streams typical of minor mergers, and associated dwarfs caught in the act of being disrupted. We intend to confirm this as part of other on-going deep surveys with the Megacam camera on the CFHT. The significance and implications of our results should also be probed using numerical simulations of galaxy evolution which may tell whether NGC~4216 has indeed had an unusually rich recent  accretion history,  if on the contrary its fellow-members have had a much  less important minor merger activity than predicted, or whether our knowledge of stellar streams and its associated time scales should be revisited.

\section*{Acknowledgments}
We wish to express our gratitude to the CFHT personnel for its dedication and tremendous help with the MegaCam observations. Many thanks to the anonymous referee for his useful and  relevant  comments.
This work is supported in part by the French Agence Nationale de la Recherche (ANR) Grant Programme Blanc VIRAGE (ANR10-BLANC-0506-01), and by the Canadian Advanced Network for Astronomical Research (CANFAR) which has been made possible by funding from CANARIE under the Network-Enabled Platforms program. J.C.M. thanks the NSF for support through grant AST-1108964. P.R.D acknowledges support from NSF grant AST-0908377. This research used the facilities of the Canadian Astronomy Data Centre operated by the National Research Council of Canada with the support of the Canadian Space Agency. The authors further acknowledge use of the NASA/IPAC Extragalactic Database (NED), which is operated by the Jet Propulsion Laboratory, California Institute of Technology, under contract with the National Aeronautics and Space Administration.

\bibliographystyle{apj}

\end{document}